# Radiative thermal runaway due to negative differential thermal emission across a solid-solid phase transition

David M. Bierman[1], Andrej Lenert[2], Mikhail A. Kats[3], You Zhou[5], Shuyan Zhang[4], Matthew De La Ossa[1], Shriram Ramanathan[6], Federico Capasso[4] and Evelyn N. Wang[1]

[1]Department of Mechanical Engineering, Massachusetts Institute of Technology, Cambridge, MA 02139, USA

[2]Department of Chemical Engineering, University of Michigan, Ann Arbor, MI 48109, USA

[3]Department of Electrical and Computer Engineering, University of Wisconsin, Madison, WI 53706, USA

[4]John A. Paulson School of Engineering and Applied Sciences, Harvard University, Cambridge, MA 02139 USA

[5]Department of Chemistry and Chemical Biology, Harvard University, Cambridge, MA 02138, USA

[6]School of Materials Engineering, Purdue University, West Lafayette, IN 47907, USA

**Abstract**

Thermal runaway occurs when a rise in system temperature results in heat generation rates exceeding dissipation rates. Here we demonstrate that thermal runaway occurs in thermal radiative systems, given a sufficient level of negative differential thermal emission. By exploiting the insulator-to-metal phase transition of vanadium dioxide, we show that a small increase in heat generation (*e.g.*, 10 nW/mm$^2$) can result in a large change in surface temperature (*e.g.*, ~35 K), as the thermal emitter switches from high emissivity to low emissivity. While thermal runaway is typically associated with catastrophic failure mechanisms, detailed understanding and control of this phenomenon may give rise to new opportunities in infrared sensing, camouflage, and rectification.

**Introduction**

Thermal runaway is a positive feedback phenomenon that occurs when a rise in system temperature promotes a further increase in its temperature. Such behavior has been observed in many different contexts, including exothermic chemical reactions [1], [2], nuclear fusion reactions [3], [4], self-heating in semiconductors [5], [6], thermal (phonon) conduction [7]–[11], fluid boiling [12], [13], and the greenhouse effect [14]. Here, we demonstrate and characterize the dynamics of thermal runaway in a radiative (photon) system. We hypothesized that thermal runaway is possible in a system where heat is dissipated from an emitter to the environment through far-field thermal radiation if the emitter exhibits sufficiently strong negative differential emission (NDE) – decreasing emission with increasing temperature. Hence, if an applied heat flux initiates strong NDE, the system temperature will continue to increase until the NDE ceases, at which point equilibrium can be re-established.

In our case, runaway was possible because of a strong dependence of the optical properties of vanadium dioxide ($VO_2$) on temperature, across its solid-solid phase transition, resulting in a modulated thermal emissivity. Vanadium oxide ($VO_2$) is a correlated-electron material that transitions from an insulator to a metal near room temperature (~340 K) [15]–[17]. This transition is accompanied by a dramatic change in optical properties in the mid-infrared spectral range [16], [18], [19]. For example, thin-films of $VO_2$ on sapphire experimentally show a strong modulation of emissivity upon phase transition [20]. Related concepts to thermal runaway such as bi-stability [21], thermal rectification [22]–[31], and thermal regulation or homeostasis [32] have been investigated in a variety of radiative systems. However, our work is the first experimental demonstration of thermal runaway in a radiative system. Thermal runaway, as demonstrated here, is also direct proof of the existence of broadband, hemispherical NDE in our geometry.

To enable applications of this phenomenon, including infrared sensing, we describe the necessary physical criteria to observe runaway (*i.e.*, the required temperature coefficient of emissivity), and the characteristic switching time of the runaway. By minimizing the thermal mass of our $VO_2$-based emitters, we show that rapid (~1 s) and reversible thermal switching is achievable.

**Criteria for NDE**

The heat flux due to thermal radiation, defined as the heat loss per unit area, is well described by the Stefan-Boltzmann law [33],

$$Q(T) = \varepsilon(T)\sigma T^4 \quad (1)$$

where $Q$ is radiative emission, $T$ is the temperature, $\varepsilon$ is the total spectrum and angle averaged emittance from a surface at a given $T$, and $\sigma$ is the Stefan-Boltzmann constant. The criterion for NDE can be determined by differentiating Equation (1),

$$dQ(T)/dT = 4\varepsilon(T)\sigma T^3 + \sigma T^4 d\varepsilon(T)/dT \quad (2)$$

For $Q$ to decrease with increasing $T$, and hence for NDE to be possible in the far field, the temperature coefficient of emittance (TCE), defined as:

$$TCE = \frac{1}{\varepsilon}\left|\frac{d\varepsilon(T)}{dT}\right| \quad (3)$$

must meet the following criterion:

$$TCE > 4/T \quad (4)$$

Note that the criterion for achieving NDE is less stringent at high temperatures. At lower temperatures, a higher value of TCE is required to observe NDE.

For $VO_2$, the phase transition occurs at ~340 K. At this temperature, the TCE must be larger than ~0.01 $K^{-1}$. We estimate, based on demonstrated $VO_2$-based emitters, that TCE is as large as 0.05 $K^{-1}$ [20], making this material a good candidate to study NDE and thermal runaway.

**Experimental Demonstration**

We grew an epitaxial $VO_2$ film (150 nm thick) on a 0.5 mm thick, polished single-crystal c-plane sapphire substrate using RF-magnetron sputtering from a $V_2O_5$ target (99.9% purity, AJA International Inc.). The substrate was held at 823 K during growth, with an RF source gun power of 125 W. We flowed 99.50 sccm of Ar and 0.5 sccm of $O_2$ as the sputtering gas mixture to keep the chamber pressure at 10 mTorr.

We then designed an experiment to study how this sample behaves when a variable amount of radiative heat flux is dissipated from its surface. The emitters were Joule heated, using a resistive heater that was sandwiched between two sapphire wafers, with the thin-film $VO_2$ deposited on both polished external surfaces (Figure 1 inset). Thermal conduction through the thermocouple and electrical leads was assumed to be negligible compared to the radiative emission loss (<5%, see Heat Transfer Model for details). Therefore, we assumed that the electrical power dissipated in the device resulted in thermal radiation either through the $VO_2$ film or the sides of the assembly.

The sample-heater-sample device was placed into a vacuum chamber near an IR-transparent viewport, which was used for visualization of the runaway process with an IR camera (sensitive to the 3-5 μm spectral range). The chamber was evacuated to 0.1 Pa, such that the heat loss by conduction from the heated surface was much smaller than that of the thermal radiation. Step-changes in voltage were applied to the heater, and the time-response of the temperature and IR signal were monitored. Quasi-steady state data was recorded when variations in the temperature measurement were within the uncertainty of the thermocouple (~0.1 K).

Figure 1 shows the heating and cooling results of the experiment. In Figure 1a, the sample temperature initially increased as the electrical power delivered to the heater increased. However, at a heater power of ~0.495 W (a total radiative heat flux of ~330 W/m$^2$), the VO$_2$ film began to change phase, resulting in NDE. Since the input power (*i.e.*, dissipated power) remained constant, we observed a thermal runaway (*i.e.*, a jump in temperature) until the structure reached thermal equilibrium with VO$_2$ in a fully metallic state. The temperature increased by ~15 °C because of the reduction in emittance. Upon cooling, we observed the reverse process wherein reducing the heat flux would induce a runaway transition back to the insulating phase; this data is also included in Figure 1a and 1b. The origin of the hysteresis is well-explained in ref. [34].

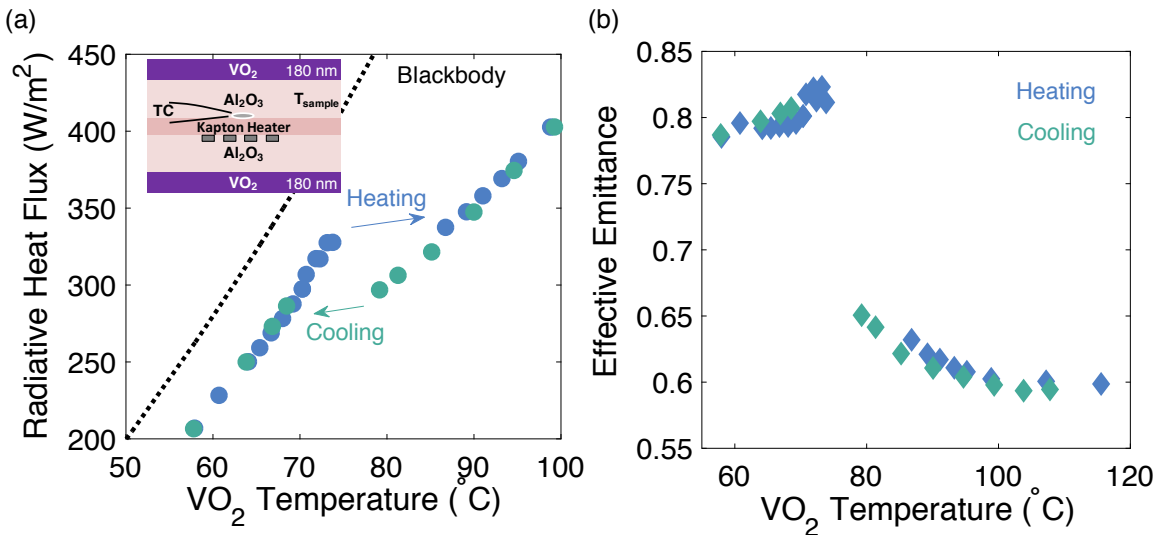

Figure 1. (a) The steady-state heating and cooling characteristics of a thin-film VO$_2$ emitter structure that undergoes a temperature-driven insulator-to-metal transition. Note that the independent variable is the radiative heat flux (vertical axis). The thermal runaway is characterized by a ~15 K jump in temperature, attributed to an abrupt reduction in the emittance of the surface. Also included is the calculated heat flux from a blackbody, for reference. Inset: Schematic of the setup where the resistive heater is embedded between two samples. (b) Effective emittance of the sample / heater assembly (measured heat flux normalized by a theoretical blackbody).

Figure 1b shows the effective emittance (*i.e.*, the radiative heat flux normalized to that of a blackbody) for the sandwich structure. This represents a weighted average of emission between the inactive (sides, supports, etc.) and active (VO$_2$ film) emission areas. The total emittance for heating exhibits an enhancement beginning at ~70 °C (due to an ultra-thin-film interference effect that is well-described in ref. [20]), before an abrupt change as the VO$_2$ film results in increasingly metallic properties over a ~20 °C window, resulting in decreasing emittance. The data in Figure 1b yield a TCE of approximately 0.014 K$^{-1}$ during the transition, and thus NDE is observable according to the criterion in Equation 4.

Figure 2a shows the system time response before, during, and after the IMT, as the input power was increased from 462 mW to 507 mW in steps of ~15 mW (red points). At temperatures far below and far above the phase transition, the observed responses (blue curve) are characteristic of a first-order dynamic system response. This response can be mathematically defined by a linear, first order differential equation with a characteristic *RC* time constant, where the thermal resistance (*R*) is due to radiation from the sample, and *C* is the heat capacity [35]. While in the insulator phase, a best-fit *RC* time constant of 4.44 minutes was determined. When the input power was stepped from 478 mW to 493 mW, the temperature rise initially had the same time constant, until the emittance began to decrease sharply at (2), reaching a maximum TCE of ~0.014 K$^{-1}$ at (3) (calculated by taking the average rate of change of emittance during the transition). The radiative thermal resistance increases as the emittance decreases, increasing the final steady state temperature and the time constant. After another increase in input power (from 493 mW to 507 mW), a second time constant was observed which is associated with the higher radiative thermal resistance of VO$_2$ in the metallic state. Here, a time constant of 7.9 minutes was extracted by fitting data after the transition (from 493 mW to 507 mW). This time constant represents an increase of ~59% relative to the insulating phase, indicating the system dynamics were heavily influenced by emission from the VO$_2$-coated areas.

Qualitative data from IR imaging in the 3-5 micron range during the heating experiment shows evidence of NDE as the cause of the thermal runaway (Figure 2b). The IR images from six representative data points in Figure 2c show that the emissive power decreases as the temperature monotonically increases. The complete IR video of this transition is available online.

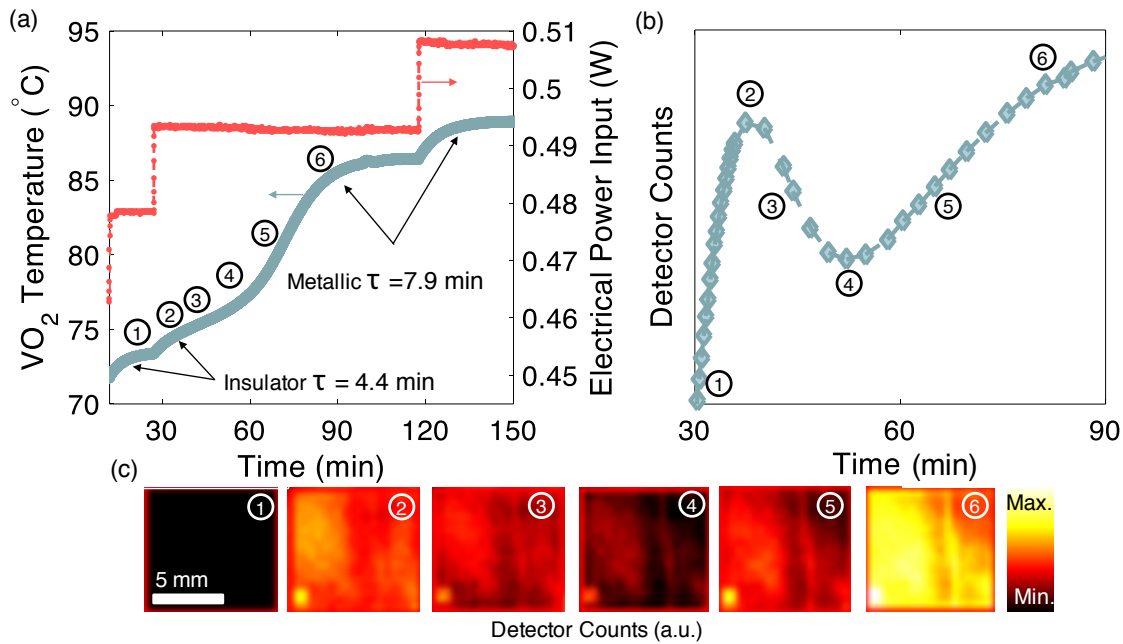

Figure 2. Dynamics of thermal runaway: (a) VO$_2$ temperature plotted as a function of time before, during, and after the transition. The secondary axis shows the applied heater power dissipated through the structure. (b) Average IR signal in the 3-5 micron range from the VO$_2$ surface during the transition. (c) IR images corresponding to the states 1-6 as denoted in the figures.

**Heat Transfer Model**

To describe the dynamics of the experiment, we constructed a lumped thermal capacitance model with a temperature dependent emittance extracted from Ref. [20]:

$$C\frac{dT}{dt} = Q_{in} - \varepsilon_{ia}A_{ia}\sigma(T^4 - T_\infty^4) - \varepsilon_{VO_2}(T)A_{VO_2}\sigma(T^4 - T_\infty^4) \quad (4)$$

where $C$ is the heat capacity of the system (extracted from the RC fit in the previous section), $T_\infty$ is the ambient temperature, $Q_{in}$ is the input heat (measured), $\varepsilon_{ia}$ is the emittance of the non-VO$_2$ surfaces (the subscripts *ia* and *VO$_2$* stand for inactive and active areas, respectively), and $A$ is the area. The emittance of inactive area on sapphire is assumed to be unity since it is covered with thick Kapton tape [36].

Figure 3a shows the results of this model, which are in good agreement with our experiments. The model accurately predicts the time constants corresponding to the insulating and metallic phases of the device, as well as the steady-state temperature. The total radiative heat loss leaving the system (Figure 3b) highlights the nature of the NDE demonstrated in our experiment. Based on the optical properties of the system, a critical heat flux (CHF, $Q_{crit}$) corresponding to ~330 W/m$^2$ (~0.49 W of input power) is defined, which when surpassed, forces the VO$_2$ to its metallic state before equilibrium can be established. The CHF is defined as the maximum possible radiative heat flux that can be supported by a surface with NDE properties without triggering thermal runaway. As described by Equation 4, this heat flow comprises both the active (from surfaces covered with VO$_2$) and inactive (from sides, leads, etc.) components (inset of Figure 3b). Over ~50% of the power is lost through the inactive area, which attenuates the NDE phenomenon since these losses have no negative-differential temperature dependence.

To investigate the limiting dynamics of thermal runaway possible given the optical properties of our thin-film VO$_2$ sample (Figures 3c, d), we used the model to study a body entirely coated by a VO$_2$ layer on top of sapphire (*i.e.*, inactive loss of 0%) with a minimum, yet realistic, heat capacity. The minimum sapphire substrate thickness that does not significantly affect the emission properties of the structure is approximately ~1 µm [20]. At $Q_{crit}$, the system equilibrates at 74 °C in an insulating phase, shown as a horizontal purple blue line in Figure 3c. However, when an additional incremental flux is supplied (*e.g.*, 1% of $Q_{crit}$), the system is forced to its metallic state through the thermal runaway process. The minimum characteristic switching time is ~1.5 seconds for a minimum temperature rise of ~35 °C. Increasing the "overpotential", or percentage above $Q_{crit}$, increases the equilibrium temperature of the device but decreases the total switching time (defined as the time elapsed between the two equilibrium states). However, the characteristic switching time ($\tau_{switch}$, defined as the time to reach (1-1/e) of the temperature difference between the two equilibrium states) is determined by the metal-phase $VO_2$ optical properties and the thermal mass, and therefore remains constant. With a heat input equal to 1% above $Q_{crit}$, the phase change is initiated, although it is delayed by ~5 seconds before the NDE begins to trigger thermal runaway (Figure 3d). With higher over-potentials (i.e., +5%), this delay can be reduced to the same characteristic switching time (~1 second), improving switching performance. Furthermore, the *RC* time constant is limited by the thermal resistance of far-field radiation. Near-field radiative transfer may reduce the switching time by several orders of magnitude by decreasing this resistance. Understanding the switching dynamics associated with radiative thermal runaway can enable the design of advanced thermal diodes, calorimeters, passive temperature control, etc. For example, sensors can benefit from this characteristic sensitivity to small incremental fluxes near the CHF to resolve small-power signals. Whereas a 1 nW change in heat input raises the temperature by ~1 °C of a 1 $mm^2$ isolated thin structure discussed above, a 1 nW additional flux near the CHF may trigger a thermal runaway and a larger change in temperature (~35 °C).

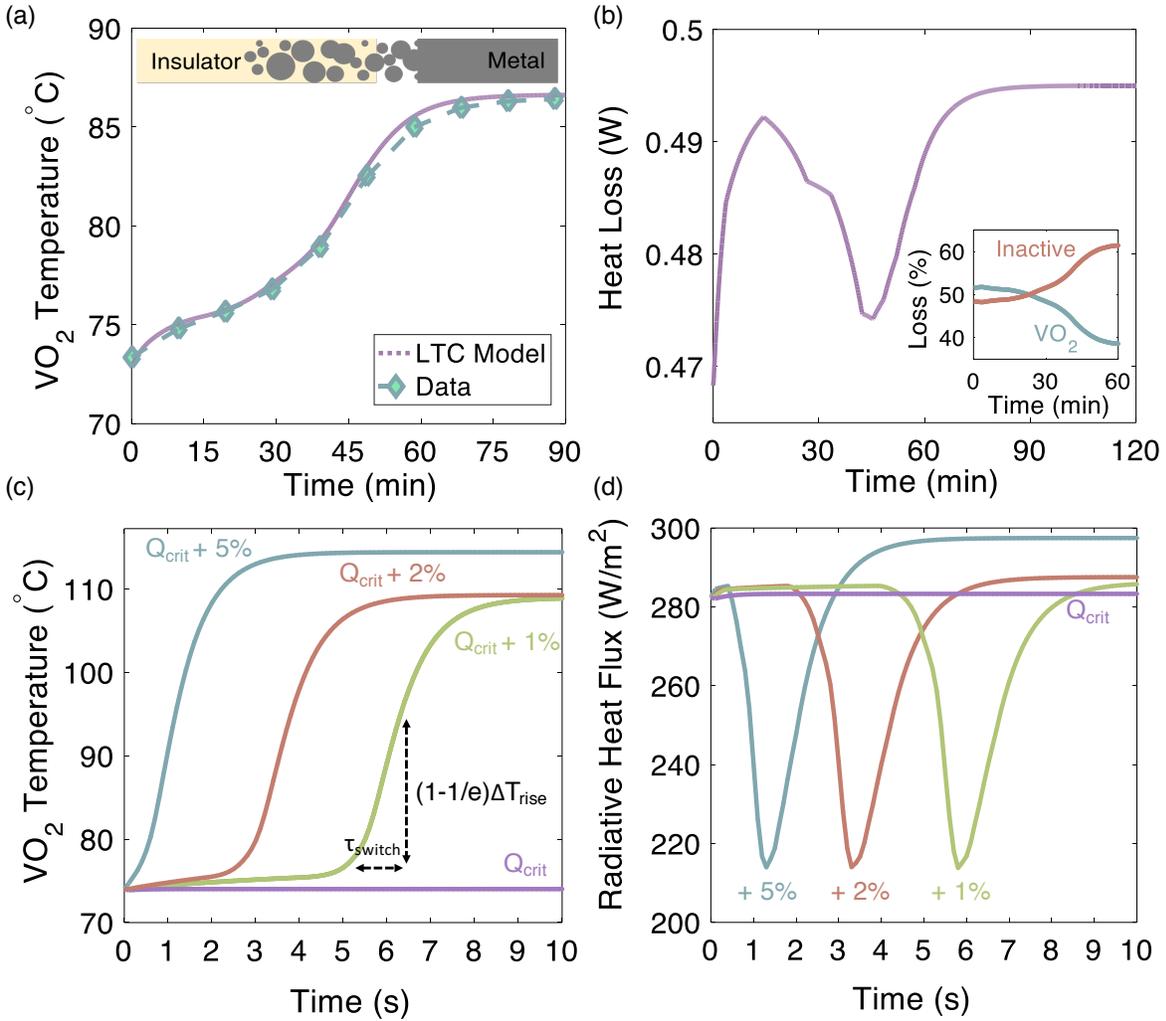

Figure 3. Simulated thermal dynamics of our experimental device and an idealized thin-film (1 μm) device, demonstrating the limits and tunability of this thermal runaway process: (a) Time-response of a lumped thermal capacitance with a temperature-dependent emittance that results from the percolation of the metallic phase. Experimental data is overlaid to show agreement. (b) Modeled transient heat loss from our device. Negative slopes of heat loss drive thermal runaway. Inset: percentage of heat loss from both the $VO_2$-covered and inactive surface areas. (c) Thermal runaway dynamics. With no over-potential above $Q''_{crit}$, the system equilibrates at 74 °C. By applying heat fluxes past this point, the onset of NDE can be modified. (d) The corresponding far-field radiative heat flux from the idealized device showing faster switching with increasing over-potential.

## Conclusions

While the widespread phenomenon of thermal runaway is often associated with catastrophic failure mechanisms, detailed understanding and control of this effect can enable new technologies for infrared sensing [37], [38], camouflage [39], and rectification [40], [41]. In a system which can only cool via thermal radiation, the temperature sensitivity of thermal emittance is the critical property that enables this phenomenon. We have shown direct evidence of NDE in a thermal-radiation system utilizing the solid-solid phase transition of vanadium dioxide, and described the dynamics of the associated thermal runaway. Its dynamics are related to both the radiative thermal resistance as well as the thermal mass of the system, leading to a high level of tunability and control of this thermal radiation phenomenon.

**Acknowledgements**

This work is supported as part of the Solid-State Solar Thermal Energy Conversion (S3TEC) Center, an Energy Frontier Research Center funded by the U.S. Department of Energy, Office of Science, Office of Basic Energy Sciences under DE-FG02-09ER46577. M. K. acknowledges support from the Office of Naval Research (Grant No. N00014-16-1-2556). S.R acknowledges support from Office of Naval Research Grant (No. N00014-16-1-2398).